# High-temperature photoluminescence reveals the inherent relations between quantum efficiency and emissivity

M. Kurtulik,[1] R. Weill,[2] A. Manor,[1] and C. Rotschild[1,2,*]

[1]Russell Berrie Nanotechnology Institute, Technion – Israel Institute of Technology, Haifa 3200003, Israel
[2]Department of Mechanical Engineering, Technion – Israel Institute of Technology, Haifa 3200003, Israel
*Corresponding author: carmelr@technion.ac.il

**Photoluminescence (PL) is a light–matter quantum interaction associated with the chemical potential μ of light formulated by the Generalized Planck's law. Without knowing the inherent temperature dependence $\mu(T)$, the Generalized Planck's law is insufficient to characterize PL(T). Recent experiments showed that PL at low temperatures conserves the emitted photon rate, accompanied by a blue-shift and transition to thermal emission at a higher temperature. Here, we theoretically study temperature-dependent PL by including phononic interactions in a detailed balance analysis. Our solution validates recent experiments and predicts important relations, including i) An inherent relation between emissivity and the quantum efficiency of a system, ii) A universal point defined by the pump and the temperature where the emission rate is fixed to any material, iii) A new phonon-induced quenching mechanism, and iv) Thermalization of the photon spectrum. These findings are relevant to and important for all photonic fields where the temperature is dominant.**

Photoluminescence (PL) conventionally involves absorption of high-energy photons followed by fast thermalization of excited electrons and subsequent emission of low-energy (red-shifted) photons. PL, first studied by Stokes [1], has been extensively researched by many others [2, 3, 4, 5, 6, 7]. Due to the complexity of many-body interactions, temperature-dependent PL at the microscopic scale is challenging to formulate; thermodynamics, however, allows it to be statistically analyzed [8, 9, 10, 11]. Treating light as ideal gas particles ascribes to PL the usual thermodynamic variables such as temperature and chemical potential [12, 13, 14]. A recent experimental study has shown [15] that PL, at elevated temperatures, conserves the emission rate up to a sharp transition to thermal emission where the rate grows exponentially. As an example, Figure 1 depicts typical temperature-dependent PL and thermal photon rates (counts per second) of $Nd^{3+}$ ions in a glass. Fig. 1a shows the temperature-dependent PL spectrum under constant 532nm laser excitation, where the spectrum is divided into two regions: 700–850nm (purple line) and 850–1000nm (green line). Fig. 1b shows the total PL rate (integrated spectrum) for each temperature (blue line). Also shown is the total integrated spectrum of only thermal emission of the same sample without optical excitation (red line), taken from Ref. 15.

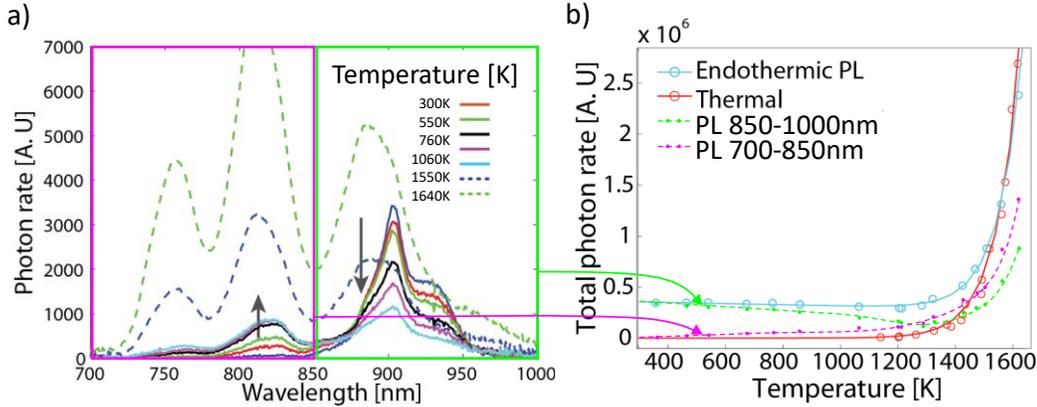

Figure 1. a) PL emission spectrum vs. temperatures showing the rate conservation associated with a blue-shift from the green region to the purple region at low temperatures (solid lines), followed by an exponential growth in intensity at any wavelength (dashed lines). b) Spectrally integrated PL (blue line) consisting of emissions from 700–85nm (purple line) and 850–1000nm (green line), Total thermal emission (without the optical excitation) is depicted by the red line.

As can be seen from Fig. 1a, an increase in temperature reduces the rate of low-energy photons at the 850–1000nm region at the cost of a rate increase of high-energy photons at 700–850nm. This blue-shift is conventionally used in optical refrigeration [16,17], and is associated with total photon rate conservation (blue line in Fig. 1b for the low temperature range). This trend continues up to a temperature above which the number of photons increases exponentially at any wavelength and reaches thermal emission at higher temperatures. Such a transition between rate conservation to exponential growth cannot be explained as the sum of a growing thermal emission and a constant PL because the second law of thermodynamics forbids the sum of two low radiation (cold) sources from resulting in a high radiation (hot) source.

To develop a model describing this transition, we begin with the description of any light source given by the Generalized Planck's formula ascribing temperature and chemical potential to PL emission [13, 14]:

$$L(h\nu, T, \mu) = \varepsilon(h\nu) \cdot \frac{2\nu^2}{c^3} \frac{1}{e^{\frac{h\nu-\mu}{k_B T}}-1} \qquad (1)$$

where $L$ is the spectral radiance (having units of watt per frequency, per solid angle, and per unit area), $T$ is the temperature, $\varepsilon$ is the emissivity, $h\nu$ is the photon energy, $K_B$ is Boltzmann's constant, and $\mu$ is the chemical potential, which is the Gibbs free energy per emitted photon and also the gap that is opened between the quasi-fermi levels under excitation in semiconductors. By its definition, the chemical potential for thermal emission is zero, $\mu = 0$. The described formalism is relevant at a specific frequency band, where the chemical potential is a constant. These relations describe any type of emission **if one knows the emissivity and chemical potential $\mu(T)$**.

We define the external quantum efficiency (EQE), which is used to describe quantum processes, as the ratio between the outgoing and incoming photon rates into the cavity, which also comprise non-absorbed incoming photons exiting the cavity. In formula (2), we define EQE at low temperatures when thermal excitation is negligible. It reflects the competition between radiative and nonradiative relaxations:

$$EQE = \frac{\#\ of\ outgoing\ photons}{\#\ of\ incoming\ photons} = \frac{L(T=0,\mu)+(1-\alpha)L_{pump}(T_p,\mu)}{L(T=0,\mu)+(1-\alpha)L_{pump}(T_p,\mu)+Q_{losses}} \quad at\ T=0K \qquad (2)$$

where $L(T=0,\mu)$ is the rate of photons emitted by a material, $\alpha$ is the absorptivity of the material, where $(1-\alpha)$ is the reflectivity of non-absorbed photons. $L_{pump}(T_p,\mu)$ is the incoming photon rate into the cavity with a brightness temperature $T_p$ and $Q_{losses}$ is the rate of absorbed photons that relax nonradiatively into heat.

In addition, we define the internal quantum efficiency (QE) to describe only material properties at low temperature:

$$QE = \frac{\text{\# of emitted photons}}{\text{\# of absorbed photons}} = \frac{L(T=0,\mu)}{L(T=0,\mu)+Q_{losses}} \quad (3)$$

Let us first consider the simple case depicted in Figure 2a. A PL body is located inside an optical cavity, in thermal contact with a heat reservoir at temperature $T$ and coupled to an optical pump source with brightness temperature $T_p$ with coupling rate $\Gamma_p$, which excites the PL body above the thermal excitation ($\mu > 0$), with outgoing coupling rate $\Gamma_{out}$. The pump's brightness temperature defines the radiance of the pump, at a specific wavelength, which is equal to a black body radiance at the same temperature, $T_p$. This is true for all solid angles and wavelengths allowed by $\Gamma_p$. The PL body in this example is represented by two energy levels (having thickness $\Delta E$) with some absorptivity $\alpha$ and an emissivity $\varepsilon$, satisfying Kirchhoff's law $\alpha = \varepsilon$, having some QE and EQE.

To understand the quasi-equilibrium temperature evolution of this two-level system, consider the temperature-dependent emission rate at a specific frequency, for the two **approximating** cases of zero-QE and unity-QE (Fig. 2b). In the former case, all the pumped photons are absorbed ($\alpha = \varepsilon = 1$) and the nonradiative relaxation rate is dominant, resulting in black body emission, regardless of the optical excitation power, and with $\mu = 0$ (Fig. 2b, black line). The latter case describes the absence of nonradiative channels where thermal excitation is negligible at every examined temperature, resulting in a conservation of the outgoing photons as the temperature changes (Fig. 2b, dashed red line). We first consider the simple case of $\Gamma_p = \Gamma_{out}$, which describes the same solid angle, spectral band and area for the incoming and outgoing photons. Under these conditions, the two lines intersect at **a point** where the thermal emission rate equals the pump rate, reflecting the equilibrium between the optical pump source and the heat reservoir $T_p = T$. This equilibrium point expresses the zero Carnot efficiency between two energy sources (reservoirs) with equal temperatures [12] that results in $\mu = 0$. The two extreme cases can also be interpreted as having QE=0 and QE=1, where QE=0 is a thermal body with emissivity $\varepsilon = \alpha$ and dominant nonradiative relaxation. The case of QE=1 describes the absence of nonradiative channels, which results in a balance between the emitted photon rate and the absorbed photon rate. Moreover, any PL emission having any EQE and QE values, between the two limiting cases, is restricted in-between these two lines (Fig. 2b, blue area), and must intersect at a single point—defining **a universal critical point.**

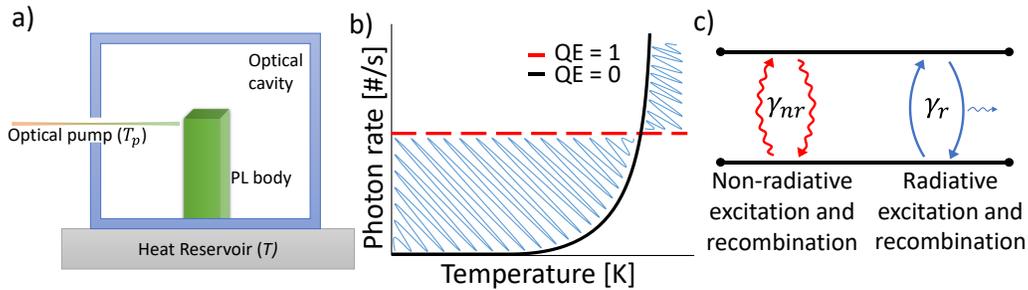

Figure 2. a) A PL body with outgoing coupling rate $\Gamma_{out}$ in contact with a heat reservoir at temperature **T** and excited with an optical pump at brightness temperature $T_p$ and coupling rate $\Gamma_p$. b) The photon rate of a PL body with EQE=1 (red line) and EQE=0 (black line). The blue area shows the limit on the emission rate for any other EQE and QE. c) A general two-level system having radiative and nonradiative interactions.

Figure 2c shows the different mechanisms involved in the detailed balance of the rates upon absorption of photons where radiative $\gamma_r$ and nonradiative $\gamma_{nr}$ rates are competing through spontaneous and stimulated processes [16]. $\gamma_{nr}$ also allows for thermal excitation, which—at thermal equilibrium (symmetric case)—compensates for nonradiative recombination. The cancelation of nonradiative processes results in emission that appears as having both 100% EQE and thermal emission. This mechanism reveals the origin of the universal critical point, as depicted in Figure 2b. As shown in the supplementary material, this is also the case for the non-symmetrical (open cavity) case where the critical point shifts to lower temperatures.

In the following, we extend this intuitive two-level system to a three-level system in a cavity described by rate equations and thereafter show the generality of the solution to any system with or without a cavity.

By following Siegman [18], we first study the case of a three-level system in a cavity (describing, for example, the 830nm–900nm emission lines of Nd$^{+3}$ [19,20] depicted in Fig. 1). Such a detailed balance considers only photonic, electronic, and phononic transitions and omits other processes associated, for example, with defects, which may cause additional temperature-dependent quenching. As such, this model describes the upper limit of temperature-dependent luminescence. Nevertheless, any additional factors can be embedded in the radiative and nonradiative rates for a specific solution. Here we assume the radiative and nonradiative rates to be temperature independent. Additional temperature-dependent values, such as the bandgap reduction with temperature rise in semiconductors, can also be implemented in the model.

Figure 3a shows the considered energy levels having a ground state and a broad excited level consisting of two closely spaced levels, with very fast nonradiative thermalization between them ($\gamma_{nr23}$). This drives towards a Boltzmann distribution of excited electron populations between the two levels, $n_2$ and $n_3$ [21] (supplementary material). Also, here, the brightness temperature $T_p$ defines the rate within the angular and spectral coupling window $\Gamma_p$. The entire system is described by:

$$\frac{dn_2}{dt} = (n_1 - n_2)B_{r12}n_{ph12} - n_2\gamma_{r12} + (n_1 - n_2)B_{nr12}n_{pn12} - n_2\gamma_{nr12} + (n_3 - n_2)B_{nr23}n_{pn23} + n_3\gamma_{nr23}, \quad (4a)$$

$$\frac{dn_3}{dt} = (n_1 - n_3)B_{r13}n_{ph13} - n_3\gamma_{r13} + (n_1 - n_3)B_{nr13}n_{pn13} - n_3\gamma_{nr13} - (n_3 - n_2)B_{nr23}n_{pn23} - n_3\gamma_{nr23}, \quad (4b)$$

$$4\pi \cdot \Delta\nu \cdot \frac{dn_{ph12}}{dt} = n_{pump12}\Gamma_p c - n_{ph12}\Gamma_{12} - (n_1 - n_2)B_{r12}n_{ph12} + n_2\gamma_{r12}, \quad (4c)$$

$$4\pi \cdot \Delta\nu \cdot \frac{dn_{ph13}}{dt} = n_{pump13}\Gamma_p c - n_{ph13}\Gamma_{13}c - (n_1 - n_3)B_{r13}n_{ph13} + n_3\gamma_{r13}, \quad (4d)$$

where $n_1, n_2, n_3$ are the electron population densities of the ground and excited states, respectively, satisfying $N = n_1 + n_2 + n_3$, where $N$ is the total density of atoms in the system; $c$ is the speed of light; $\gamma_{r12}, \gamma_{r13}$ are radiative spontaneous rates, $\gamma_{nr12}, \gamma_{nr13}$ are nonradiative spontaneous rates, from both excited levels to the ground state, with units of [1/s]; $\gamma_{nr23} \gg \gamma_{r12}, \gamma_{r13}, \gamma_{nr12}, \gamma_{nr13}$ is the nonradiative rate between excited energy levels; $\Gamma_p, \Gamma_{12}, \Gamma_{13}$ are the coupling rates in and out of the cavity, respectively; $n_{ph12}, n_{ph13}$ are the radiation field densities inside the cavity, having units of $\left[\frac{\#}{\Delta\nu \cdot Sr \cdot m^3}\right]$; $n_{pump12} = DoS_{ph}\left[\exp\left(\frac{E_{12}}{kT_p}\right) - 1\right]^{-1}$, $n_{pump13} = DoS_{ph}\left[\exp\left(\frac{E_{13}}{kT_p}\right) - 1\right]^{-1}$ are the optical fields induced by the pump having a black-body distribution according to the brightness temperature $T_p$; and $B_{r12} = \gamma_{r12}/DoS_{ph}(E_{12})$, $B_{r13} = \gamma_{r13}/DoS_{ph}(E_{13})$, and $B_{nr12} = \gamma_{nr12}/DoS_{pn}(E_{12})$, $B_{nr13} = \gamma_{nr13}/DoS_{pn}(E_{13})$ are the Einstein coefficients for the stimulated absorption or emission rates for both radiative and nonradiative processes [8]. Under fast thermalization, phonons obey the equilibrium distribution, which is given by $n_{pn12} = DoS_{pn}(E_{12})\left[\exp\left(\frac{E_{12}}{kT}\right) - 1\right]^{-1}$, $n_{pn13} = DoS_{pn}(E_{13})\left[\exp\left(\frac{E_{13}}{kT}\right) - 1\right]^{-1}$ and $n_{pn23} = DoS_{pn}(E_{23})\left[\exp\left(\frac{E_{23}}{kT}\right) - 1\right]^{-1}$ [12]. In this formalism, $DoS_{pn}$ and $DoS_{ph}$ are the corresponding densities of states (DoS) for the phonons and photons, respectively.

The solutions of equations (3a–3d) for the photon rate (radiative transitions to the ground state), given by $n_{ph12}\Gamma_{12}$ and $n_{ph13}\Gamma_{13}$, for equal incoming and output coupling rates; $\Gamma_{12} = \Gamma_{13} = \Gamma_p = \Gamma$, under optical excitation $n_{pump12}\Gamma_p$ and $n_{pump13}\Gamma_p$ at $T_p = 1000K$ for various QEs of 0, 0.5, and 1, are depicted in Figure 3b. These three cases all have the same absorptivity (as well as emissivity), $\alpha(\gamma_r) = \varepsilon(\gamma_r)$, and different losses due to different $\gamma_{nr}$ values.

The red line describes both the absorbed pump rate and the emission of QE=1. The black line in Figure 3b describes the emission of a thermal body with the same emissivity $\varepsilon(\gamma_r)$, when setting QE=0, describing the case where $\gamma_{nr12}, \gamma_{nr12} \gg \gamma_{r12}, \gamma_{r13}$. Under this regime, nearly all the absorbed photons recombine nonradiatively, which results in thermal emission that increases exponentially until it reaches the QE=1 line at the universal point, $T_c = T_p$. The blue line describes the PL emission of QE=0.5. At 0K, half of the absorbed photons are lost due to nonradiative recombination $\gamma_{nr}$. At a low (non-zero) temperature range, the total photon rate (sum of $n_{ph12}\Gamma_{12}$ and $n_{ph12}\Gamma_{13}$) is quasi-conserved, accompanied by the blue-shift of the spectrum–emitted photon rate $n_{ph12}\Gamma_{12}$ that decreases while $n_{ph12}\Gamma_{12}$ increases, as the temperature increases. The ratio between these emissions is given by the Boltzmann distribution as long as $\gamma_{nr23} \gg \gamma_{r12}, \gamma_{r13}, \gamma_{nr12}, \gamma_{nr13}$ [21]. A further increase in temperature leads to an increase in the photon rate until it reaches the cross between the QE=1 case (red line) and the QE=0 case (black line) at the universal point $T_c = T_p$, and then continues to rise above it.

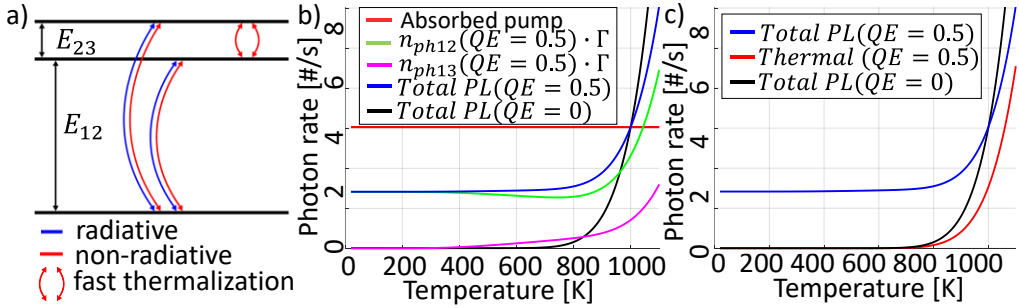

Figure 3. a) A three-level system with fast nonradiative thermalization between upper energy levels. b) PL emission for three different QE=0, 0.5 and 1 systems (black line, blue line and red line, respectively), optically excited by a pump at brightness temperature $T_p = 1000K$. The green and purple lines show individual emissions from lower and higher excited energy levels for the QE=0.5 case. c) Above the critical temperature $T_c < T$, the PL emission (blue line) is bounded by a thermal body when QE=0 (black line) and by the thermal emission for the specific QE (QE=0.5, red line).

For a $n_{pump} = 0$, one gets a thermal emission for any QE. Figure 3c shows the PL line for QE=0.5 (blue line) in addition to its corresponding thermal emission (red line). For comparison, we also show thermal emission for the case of QE=0 (black line). As evident, the two emission lines, PL and thermal for QE=0.5, merge together at high temperatures.

This general solution, is as far as we know, the first theoretical explanation for the experimentally observed transition from the rate-conservation region accompanied by a blue-shift to thermal emission, where the photon rate increases at any wavelength, as shown in Figure 1. The thermal emission for QE>0 is reduced, compared to the thermal emission of QE=0, due to a lower value of $\gamma_{nr}$ compared to $\gamma_r$. As evident, the PL emission beyond the universal point is restricted to remaining between the QE=0 line and the thermal curve ($n_{pump} = 0$) for the same QE.

The ratio between the thermal curve of a specific material (QE>0) and the QE=0 emission curve is a temperature-independent and QE-dependent constant we name $\varepsilon_{QE}(\gamma_{r12}, \gamma_{r13}, \gamma_{nr12}, \gamma_{nr13})$ and it is given by: $\varepsilon_{QE}(\gamma_{r12}, \gamma_{r13}, \gamma_{nr12}, \gamma_{nr13}) = 1 - QE$ (see the analytical solution for the two-level system in the supplementary material).

Figure 4a depicts this linear relation between $\varepsilon_{QE}$ and QE for a three-level system. The conventional emissivity can be expressed as:

$$\varepsilon_{QE>0} = \varepsilon_{QE=0}(DoS, \gamma_{r12}, \gamma_{r13}) \times \varepsilon_{QE}(\gamma_{r12}, \gamma_{r13}, \gamma_{nr12}, \gamma_{nr13}) = \alpha \times (1 - QE), \quad (5)$$

Eq. 5 reflects the non-equilibrium conditions, where a material thermally emits toward empty space (absence of pump) at a rate of:

$$L(h\nu, T) = \Gamma \times \alpha \times (1 - QE) \times \frac{2\nu^2}{c^3} \frac{1}{e^{\frac{h\nu}{k_B T}} - 1}, \quad (6)$$

This formalism is in line with Kirchhoff's law, when placing such a PL emitter in equilibrium with a black body. The black body pumps the PL emitter and the system relaxes at the critical point. This is, as far as we know, the first indication of an inherent dependency between QE and the total emissivity. The effect of closing the cavity is depicted in Figure 4b, which shows the QE for various values of $\gamma_{r12}, \gamma_{r13} = \gamma_r$ and $\gamma_{nr12}, \gamma_{nr13} = \gamma_{nr}$ and different coupling values of $\Gamma_{12} = \Gamma_{13} = \Gamma$.

The QE approaches zero when the coupling rate $\Gamma$ is reduced, while the emissivity approaches unity. In this model, $\gamma_r$ and $\gamma_{nr}$ are not $\Gamma$ dependant. This is the case when cavity dimensions are much larger than the emission wavelength and the Purcell factor converges to unity [22].

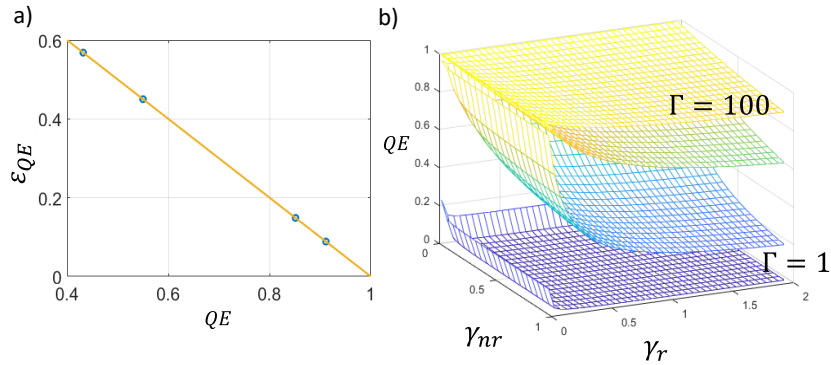

Figure 4. a) Linear relation between the emissivity $\varepsilon_{QE}$ and QE. b) QE vs. $\gamma_{nr}$ and $\gamma_r$ for different coupling rates $\Gamma$.

Thus far, we considered a pump, which is fully defined by the brightness temperature at any wavelength and the solid angle allowed by the coupling $\Gamma_{12} = \Gamma_{13} = \Gamma$. The solution can describe qualitatively the experimental results from Figure 1, where the system is pumped at the 4th level, by having different $\gamma_{r12}$ and $\gamma_{r13}$. Figure 5 shows the solution when we allow Gaussian broadening on each discreet level. As evident, the solution resembles the experimental results.

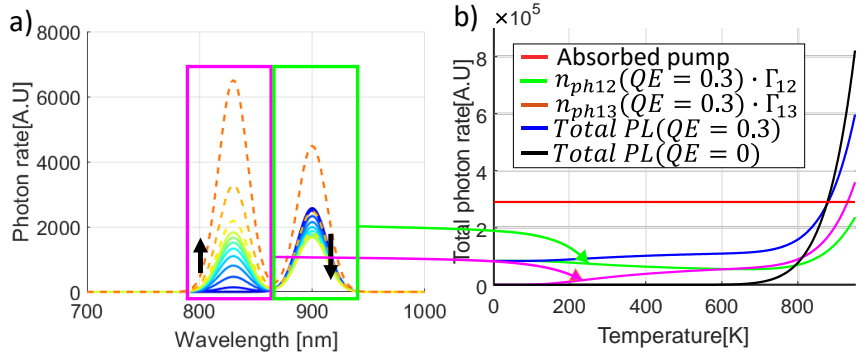

Figure 5. Simulation of the rate equations, qualitatively explaining Figure 1's experimental results. a) The temperature-dependent PL spectrum exhibits blue-shift and transition to thermal emission. b) Spectrally integrated PL for QE=0.3, plotted for comparison with QE=0.

If one looks carefully at the PL total photon rate in Figure 1, one can notice that with the increasing temperature, the total photo rate slightly decreases before it sharply increases. This can also be seen in the model in Figure 6 for the case where $\gamma_{nr13} > \gamma_{nr12}$. At high temperatures (within the quasi-conservation region), electrons promoted to level-3 nonradiatively decay at an enhanced rate.

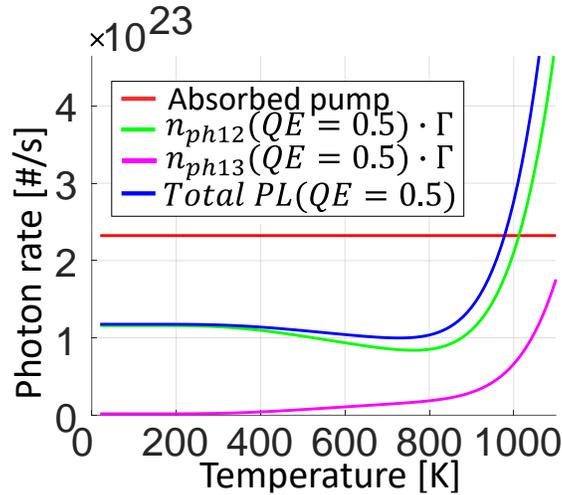

Figure 6. Simulation result of total photon rate for QE=0.5, when $\gamma_{nr13} > \gamma_{nr12}$, showing the rate decrease at high temperatures (within the quasi-conservation region).

Interestingly, our model goes beyond the specific experiment described in Figure 1 and predicts a new temperature-induced quenching for a system having a broad ground state. We discuss this phenomenon in the supplementary material.

Finally, our model also supports recent experiments and theory claiming the thermalization of the PL spectrum and a spectral emissivity approaching unity when closing the cavity (minimizing Γ), and the spectrum evolves into a Boltzmann distribution [23, 24] (see supplementary material).

In summary, we developed a model for temperature-dependent luminescence using a detailed balanced formalism at high temperatures, where thermal excitation is comparable to photonic excitation. Our results support the experimental observations of photoluminescence at elevated temperatures, exhibiting a blue-shift of the spectrum while the photon rate is quasi-conserved and it transitions to thermal emission. We also show the existence of a universal point where the emission rate of any QE, under a similar absorbed pump and temperature, is fixed. More generally, our model shows an inherent dependency between QE and emissivity, which is important in lighting and energy harvesting systems as well as in any field of radiation where the evaluation of the limit of radiation is critical.

**Funding.** European Union's Seventh Framework Program (H2020/2014-2020]) under grant agreement n° 638133-ERC-ThforPV

**Disclosures.** The authors declare no conflicts of interest.

**Data availability.** The data that support the findings of this study are available from the corresponding author upon reasonable request.

See Supplement 1 for supporting content.

# Supplementary material
## Two-level system
Here, we present a theoretical formalism for the excited electron and photon populations inside the cavity:

$$\frac{dn_2}{dt} = (n_1 - n_2)B_{r12}n_{ph12} - n_2\gamma_r + (n_1 - n_2)B_{nr12}n_{pn12} - n_2\gamma_{nr}, \tag{S.1a}$$

$$4\pi \cdot \Delta\nu \cdot \frac{dn_{ph12}}{dt} = n_{pump}\Gamma_p - n_{ph12}\Gamma_{out} - (n_1 - n_2)B_{r12}n_{ph12} + n_2\gamma_r, \tag{S.1b}$$

Using the same notation as for the three-level case in the paper: $n_1$ and $n_2$ are the electron population densities of the ground and excited states ($N = n_1 + n_2$); $\gamma_r$ and $\gamma_{nr}$ are radiative and nonradiative spontaneous rates; $B_{r12} = \gamma_r/DoS_{ph}$ and $B_{nr12} = \gamma_{nr}/DoS_{pn}$ are radiative and nonradiative stimulated rates; $\Gamma_p$ and $\Gamma_{out}$ are coupling rates in and out of the cavity, respectively; $n_{ph}$ and $n_{pn}$ are the field densities for photons and for phonons, respectively, where $n_{pump} = DoS_{ph}\left[\exp\left(\frac{E_g}{kT_p}\right) - 1\right]^{-1}$ and $n_{pn} = DoS_{pn}\left[\exp\left(\frac{E_g}{kT}\right) - 1\right]^{-1}$; and $DoS$ are the density of states (for photons or phonons).

In the symmetrical case, when $\Gamma_{out} = \Gamma_p$, Figure S.1a shows the PL emission for various QEs (0 – black line, 0.5 – blue line and 1 – red line). The critical point occurs at $T_c = T_p$, which is the equilibrium point between the system and the optical source $n_{pump}$. Figure S.1b shows the same QEs (0, 0.5 and 1) for the nonsymmetrical case, when $\Gamma_{out} > \Gamma_p$. This reflects a situation when the PL emission is emitted in a wider solid angle compared to the solid angle of the optical excitation. The critical point occurs at $T_c = \frac{E_g}{k}\left(\ln\left(\frac{\Gamma_{out}}{\Gamma_p}\left[e^{\frac{E_g}{kT_p}} - 1\right] + 1\right)\right)^{-1}$. One can think about this case from the point of view of a normalized excitation source: the system is excited by an effective pump $n_{pumpEff} = DoS_{ph}\left[\exp\left(\frac{E_g}{kT_c}\right) - 1\right]^{-1} = n_{pump}\Gamma_p/\Gamma_{out}$ with brightness temperature $T_c$.

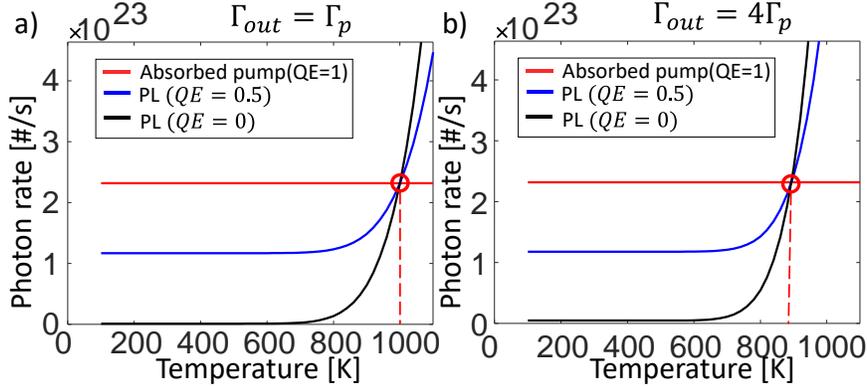

Figure S.1. a) PL emission for three different QE=0, 0.5 and 1 systems (black line, blue line and red line), optically excited by a pump at brightness temperature $T_p = 1000K$ (red line, which is also the PL emission for QE=1), with symmetrical coupling rates $\Gamma_{out} = \Gamma_p$; the critical temperature is $T_c = T_p = 1000K$. b) PL emission for the same QE=0, 0.5 and 1 systems—the nonsymmetrical case, $\Gamma_{out} = 4\Gamma_p$; the critical temperature is $T_c = 840K < T_p$.

## Two-level system, universal point
Here, we show analytically the existence of the critical point for the two-level system. For $N \gg n_2$, which is correct for excitation much lower than the population inversion, equations S.1a–b at steady-state become:

$$0 = NB_{r12}n_{ph12} - n_2\gamma_r + NB_{nr12}n_{pn12} - n_2\gamma_{nr}, \tag{S.2a}$$

$$0 = n_{pump}\Gamma_p - n_{ph12}\Gamma_{out} - NB_{r12}n_{ph12} + n_2\gamma_r, \tag{S.2b}$$

Equation S.2a is solved for $n_2$, and we substitute it into equation S.2b to get the rate of photons going out of the cavity:

$$n_{ph12}\Gamma_{out} = \Gamma_{out}\frac{n_{pump}\Gamma_p + NB_{nr12}n_{pn12}\frac{\gamma_r}{\gamma_r+\gamma_{nr}}}{\Gamma_{out} + NB_{r12}(1-\frac{\gamma_r}{\gamma_r+\gamma_{nr}})}, \tag{S.3}$$

Using $n_{pump} = DoS_{ph} \left[\exp\left(\frac{E_g}{kT_p}\right) - 1\right]^{-1} = \frac{\gamma_r}{B_{r12}} f(T_p)$, where $f(T) = \left[\exp\left(\frac{E_g}{kT}\right) - 1\right]^{-1}$ and the term for nonradiative interactions $B_{nr12} n_{pn12} = f(T) \gamma_{nr}$, we get:

$$n_{ph12}\Gamma_{out} = \Gamma_{out} \frac{\frac{\gamma_r}{B_{r12}} f(T_p)\Gamma_p + N f(T)\gamma_{nr} \frac{\gamma_r}{\gamma_r + \gamma_{nr}}}{\Gamma_{out} + NB_{r12}\left(1 - \frac{\gamma_r}{\gamma_r + \gamma_{nr}}\right)} = \frac{\gamma_r}{B_{r12}} f(T_p)\Gamma_p \frac{\Gamma_{out} + \frac{\Gamma_{out}}{\Gamma_p} \frac{f(T)}{f(T_p)} NB_{r12} \frac{\gamma_{nr}}{\gamma_r + \gamma_{nr}}}{\Gamma_{out} + NB_{r12} \frac{\gamma_{nr}}{\gamma_r + \gamma_{nr}}}$$

$$n_{ph12}\Gamma_{out} = n_{pump}\Gamma_p \frac{\Gamma_{out} + \frac{\Gamma_{out} f(T)}{\Gamma_p f(T_p)} NB_{r12} \frac{\gamma_{nr}}{\gamma_r + \gamma_{nr}}}{\Gamma_{out} + NB_{r12} \frac{\gamma_{nr}}{\gamma_r + \gamma_{nr}}}, \tag{S.4}$$

For $n_{ph12}\Gamma_{out} = n_{pump}\Gamma_p$, at $T = T_c$, the ratio has to be one:

$$\frac{\Gamma_{out} + \frac{\Gamma_{out} f(T_c)}{\Gamma_p f(T_p)} NB_{r12} \frac{\gamma_{nr}}{\gamma_r + \gamma_{nr}}}{\Gamma_{out} + NB_{r12} \frac{\gamma_{nr}}{\gamma_r + \gamma_{nr}}} = 1, \tag{S.5}$$

and this is only if $\frac{\Gamma_{out}}{\Gamma_p} \frac{f(T_c)}{f(T_p)} = 1$. In the symmetrical case when $\Gamma_p = \Gamma_{out}$, this leads to $T_c = T_p$. In the nonsymmetrical case, the critical temperature is lower, fulfilling the equation $f(T_c) = \frac{\Gamma_p}{\Gamma_{out}} f(T_p)$.

**Two-level system, emissivity**

The external quantum efficiency (EQE) for a two-level system at $0K$ can be written as:

$$EQE = \frac{n_{ph12}\Gamma_{out}}{n_{pump}\Gamma_p} = \frac{\Gamma_{out}}{\Gamma_{out} + NB_{r12}(1 - \frac{\gamma_r}{\gamma_r + \gamma_{nr}})}, \tag{S.6}$$

For the QE=0 case where $\gamma_{nr} \gg \gamma_r$, equation 5 becomes:

$$EQE(\gamma_{nr} \gg \gamma_r) = \frac{1}{1 + \frac{NB_{r12}}{\Gamma_{out}}}, \tag{S.7}$$

EQE>0 while QE=0 can occur due to the non-absorbed pump reflected back from the cavity. Thus, the absorptivity of the PL material is:

$$\alpha = 1 - EQE(\gamma_{nr} \gg \gamma_r) = \frac{1}{\frac{\Gamma_{out}}{NB_{r12}} + 1}, \tag{S.8}$$

The internal quantum efficiency (QE) can be written as:

$$QE = \frac{n_{ph12}\Gamma_{out}c - (1-\alpha)n_{pump}\Gamma_p c}{\alpha n_{pump}\Gamma_p c} = \frac{EQE}{\alpha} - \frac{(1-\alpha)}{\alpha} = \frac{\frac{\gamma_r}{\gamma_r + \gamma_{nr}}}{1 + \frac{NB_{r12}}{\Gamma_{out}}\left(1 - \frac{\gamma_r}{\gamma_r + \gamma_{nr}}\right)}, \tag{S.9}$$

In the model, we assume that the internal material parameters such as $\gamma_r$ and $\gamma_{nr}$ and $DoS_{ph}, DoS_{pn}$ are temperature independent. Therefore, the emissivity for a given $\gamma_r$ and $\gamma_{nr}$, and thus $QE$, is temperature independent. It can be analytically shown, from equation S.3, using relation S.9, that the thermal emission of this system, when $n_{pump} = 0$, is given by:

$$n_{ph12}\Gamma_{out} = \Gamma_{out} c f(T) DoS_{ph} \alpha \cdot (1 - QE), \tag{S.10}$$

Comparing equations (S.10) with Planck's radiation law, $L(T) = \varepsilon_{QE>0} \cdot DoS_{ph} f(T)$, we come to the conclusion that the emissivity $\varepsilon_{QE>0} = \alpha \cdot (1 - QE)$.

Figure S.2 shows the PL of $QE=o.5$ (blue line) and its thermal emission (purple line) is in line with Planck's radiation law with $\alpha = 1$ and the emissivity given by $\varepsilon_{QE>0}$ (black circles).

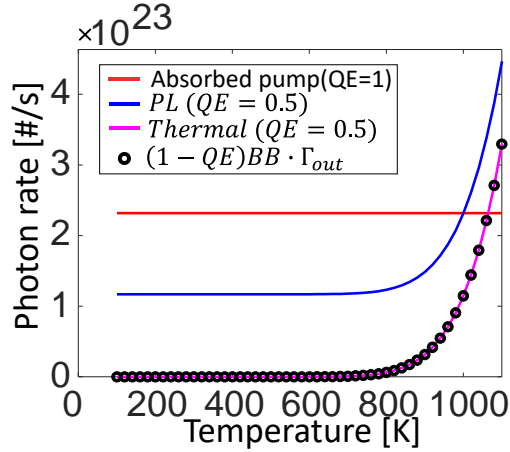

Figure S.2. PL emission (blue) of *QE=0.5* and its thermal emission (purple) fitted with black body emission ($\alpha = 1$) with the emissivity $\varepsilon_{QE>0} = \alpha \cdot (1 - QE)$ (black circles).

## Three-level system
### Fast evolution of the excited electron populations to the Boltzmann distribution
The electron populations in the three-level system are described by equations 4a and 4b in the main body of the paper. For more convenience, we write these equations here as well:

$$\frac{dn_2}{dt} = (n_1 - n_2)B_{r12}n_{ph12} - n_2\gamma_{r12} + (n_1 - n_2)B_{nr12}n_{pn12} - n_2\gamma_{nr12} + (n_3 - n_2)B_{nr23}n_{pn23} + n_3\gamma_{nr23}, \text{ (S.11a)}$$

$$\frac{dn_3}{dt} = (n_1 - n_3)B_{r13}n_{ph13} - n_3\gamma_{r13} + (n_1 - n_3)B_{nr13}n_{pn13} - n_3\gamma_{nr13} - (n_3 - n_2)B_{nr23}n_{pn23} - n_3\gamma_{nr23}, \text{ (S.11b)}$$

The last two terms in both equations are nonradiative interactions between $n_2$ and $n_3$. $\gamma_{nr23}$ is the spontaneous nonradiative rate and $B_{nr23} = \frac{\gamma_{nr23}}{DoS_{pn}}$ is the stimulated nonradiative rate. The electronic levels interact with the phonon field, given by the equilibrium distribution $n_{pn23} = DoS_{pn}\left[\exp\left(\frac{E_{23}}{kT}\right) - 1\right]^{-1}$. Given this, we can write these two last terms as:

$$(n_3 - n_2)B_{nr23}n_{pn23} + n_3\gamma_{nr23} = \gamma_{nr23}\left(\left[\exp\left(\frac{\Delta E}{kT}\right) - 1\right]^{-1}[n_3 - n_2] + n_3\right)$$

$$= n_3\gamma_{nr23}\left(\left[\exp\left(\frac{\Delta E}{kT}\right) - 1\right]^{-1}\left[1 - \frac{n_2}{n_3}\right] + 1\right) = n_3\gamma_{nr23}\left(1 - \frac{\left[\frac{n_2}{n_3} - 1\right]}{\left[\exp\left(\frac{\Delta E}{kT}\right) - 1\right]}\right), \quad \text{(S.12)}$$

In the case when $\gamma_{nr23}$ is dominant, at steady-state where $\frac{dn_2}{dt} = \frac{dn_3}{dt} = 0$, which can be satisfied when these terms approach zero leading to a Boltzmann distribution, we have:

$$n_3\gamma_{nr23}\left(1 - \frac{\left[\frac{n_2}{n_3} - 1\right]}{\left[\exp\left(\frac{\Delta E}{kT}\right) - 1\right]}\right) \approx 0 \rightarrow \frac{n_2}{n_3} \approx \exp\left(\frac{\Delta E}{kT}\right), \quad \text{(S.13)}$$

### Three-level system, temperature induced quenching
Figure S.3a depicts a system having a broad ground state represented by two thermally coupled energy levels. Figure 3b shows a decrease in the photon rate at temperatures lower than the critical temperature due to a stimulated nonradiative interaction. At low temperatures, when phononic excitation is negligible, $n_2$ is empty, while spontaneous radiative and nonradiative recombinations from the $n_3$-level are dominant. Increasing the temperature slightly, increases the population of $n_2$, which allows reabsorption of photons with energy $E_{23}$. Some of these photons will be nonradiatively thermalized and lost. This results in a decrease of the total photon emission. A further temperature rise leads to thermal excitation of electrons from the ground state and again to an increase in total emission.

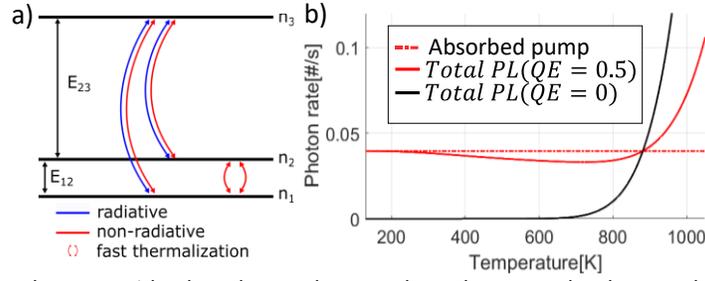

Figure S.3. a) The three-level system with a broad ground state, where the energy levels $n_1$ and $n_2$ are thermally coupled. b) A numerical simulation of the given system with a total photon rate decreasing below the critical point.

## Thermalization of the PL spectrum

As was experimentally and theoretically shown, any emission spectrum in an open cavity thermalizes in a closed cavity, where reabsorption and re-emission shifts the spectrum towards a black body [1,2]. To show this thermalization of the PL spectrum, we use a two-level system that has a broad photon spectrum $n_{ph}(\nu)$. The equations for the excited population $n_2$ and the photons $n_{ph}(\nu)$ are given by:

$$\frac{dn_2}{dt} = \int d\nu [n_1\sigma_{12}(\nu) - n_2\sigma_{21}(\nu)]n_{ph}(\nu) - n_2 \int d\nu \sigma_{21}(\nu)\frac{8\pi\nu^2}{c^2} + (n_1 - n_2)B_{nr}n_{pn} - n_2\gamma_{nr}, \quad (S.14a)$$

$$4\pi \cdot \Delta\nu \cdot \frac{dn_{ph}(\nu)}{dt} = -n_{ph}(\nu)\Gamma c - [n_1\sigma_{12}(\nu) - n_2\sigma_{21}(\nu)]n_{ph}(\nu) + n_2\sigma_{21}(\nu)\frac{8\pi\nu^2}{c^2}, \quad (S.14b)$$

where $\sigma_{12}(\nu)$ and $\sigma_{21}(\nu)$ are the absorption and emission cross-sections. In this model, we use Erbium absorption and emission cross-sections. Figure 4 depicts the evolution of the spectra for various $\Gamma$. As depicted, when the optical cavity is closing, $\Gamma$ becomes smaller, and the Erbium spectrum shifts towards a black body spectrum that is exponentially growing.

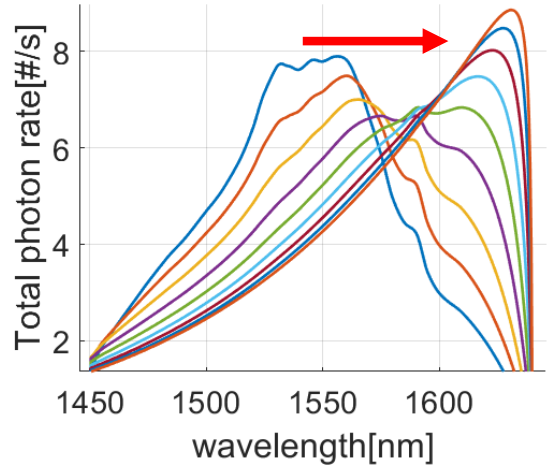

Figure 4. Thermalization of an Erbium PL spectrum